\documentclass[aps,prd,preprint,groupedaddress,showpacs,amsmath]{revtex4}

\usepackage{graphicx}
\usepackage{color}

\def\twobg{2\beta+\gamma}
\def\sinbg{\sin(2\beta + \gamma)}
\def\cosbg{\cos(2\beta + \gamma)}

\def\sss{\scriptscriptstyle}

\def\barp{{\raise.35ex\hbox {${\sss (}$}}---{\raise.35ex\hbox{${\sss )}$}}}
\def\bdbarp{\hbox{$B_d$\kern-1.4em\raise1.4ex\hbox{\barp}}}
\def\bsbarp{\hbox{$B_s$\kern-1.4em\raise1.4ex\hbox{\barp}}}
\def\barpd{{\raise.35ex\hbox {${\sss (}$}}--{\raise.35ex\hbox{${\sss )}$}}}
\def\dbarp{\hbox{$D^{*0}$\kern-1.6em\raise1.5ex\hbox{\barpd}}}
\def\kbarp{\hbox{$K^{*0}$\kern-1.6em\raise1.5ex\hbox{\barpd}}}

\def\roughly#1{\mathrel{\raise.3ex\hbox{$#1$\kern-.75em\lower1ex\hbox{$\sim$}}}}

\def\Re{{\rm  Re}}
\def\Im{{\rm Im}}
\def\btou{b\to u \overline c d}
\def\btoc{b\to c \overline u d}
\def\B{{\cal B}}
\def\fbar{\overline f}
\def\Bz{B^0}
\def\Bzb{{\overline B^0}}
\def\Sex{S_{\pi/2}}
\def\Spi{S'_\pi}
\def\Spm{S'_\pm}

\begin{document}

\title{Improved measurement of $2 \beta + \gamma$}

\author{Nita Sinha}
\author{Rahul Sinha}
\affiliation{Institute of Mathematical Sciences, Taramani, Chennai 600113, India}

\author{Abner Soffer}
\affiliation{Department of Physics, 
        Colorado State University, Fort Collins, Colorado 80523, USA.}

\date{\today}

\begin{abstract}
We propose to measure the Cabibbo-Kobayashi-Maskawa parameter $\twobg$
using $B^0$ decays involving several intermediate states, and describe
a general formalism that applies to a broad class of decays. The
main advantage of this method is that the ratios between the
interfering amplitudes can be measured without requiring external
input. In addition, discrete ambiguities are resolved.

\end{abstract}


\maketitle

\section{Introduction}

CP violation is one of the most important topics in current particle
physics research. In the standard model, CP violation arises due to a
single complex phase in the Cabibbo-Kobayashi-Maskawa
matrix $V$~\cite{ref:CKM}.  A major goal of $B$ meson physics is 
to measure the angles and sides of the CKM unitarity
triangle. Theoretically clean measurement methods are crucial for
obtaining these parameters accurately. The BaBar~\cite{ref:Babar-sin2b}
and Belle~\cite{ref:Belle-sin2b} measurements of the parameter $\sin(2\beta)$,
where $\beta = \arg{\left(- V_{cd} V_{cb}^\ast/ V_{td} V_{tb}^\ast\right)}$,
confirm the standard model to within the precision of the experiments,
and increased precision is expected in the coming years.

Crucial studies of the CKM mechanism and constraints on 
new physics can be obtained by measuring
the CKM angle 
$\gamma = \arg{\left(- V_{ud} V_{ub}^\ast/ V_{cd} V_{cb}^\ast\right)}$.
The greatest challenges presented by these measurements is
that they require very large data samples and are subject to discrete
ambiguities. It is therefore important to use every possible mode and
method for measuring $\gamma$, and to devise methods that help resolve
the ambiguities.

An important class of measurements
makes use of decays such as $B\to D^{-} \pi^+$ to measure
$\twobg$. Proposed initially by Dunietz~\cite{ref:dunietz}, the first
attempts to measure time-dependent CP asymmetries proportional to
$\sinbg$ and $\cosbg$ have been conducted by
BaBar~\cite{ref:babar-sinbg} and Belle~\cite{ref:belle-sinbg} using
the modes $B\to D^{(*)-} \pi^+$ and $B\to D^{-} \rho^+$.  
While these measurements are currently statistically limited, their
precision will become significant as more data are accumulated. At
that stage, the greatest difficulty in extracting $\twobg$ from these
results will be the lack of precise knowledge of the ratio between the
interfering amplitudes, defined as 
$r \equiv |A(\overline B^0\to D^{(*)-}h^+) / A(B^0\to D^{(*)-}h^+)|$,
where $h^+$ indicates the light hadron $\pi^+$ or $\rho^+$.

In principle, $r$ may be obtained from the difference between the
magnitudes of two terms with different time dependences in the decay
rate. The relevant terms are $(1 + r^2)$ and $(1 - r^2) \cos \Delta mt$,
where $\Delta m$ is the $\Bz - \Bzb$ oscillation frequency.  
However, with $r \sim {\cal O}(1-2\%)$, extracting it from the ${\cal O}(r^2)$ 
difference 
between these ${\cal O}(1)$ terms requires prohibitively large data
sets.
Thus, the time-dependent measurement has negligible
sensitivity to the value of $r$, which must therefore be 
obtained by assuming factorization and 
SU(3) symmetry to make use of the ratio of branching fractions 
$\B(\overline B^0\to D_s^{(*)-}h^+) / \B(B^0\to D^{(*)-}h^+)$. 
This approximation ignores the contribution of annihilation diagrams
and some SU(3) breaking effects, and is taken to have a theoretical
error of roughly 30\%~\cite{ref:babar-sinbg}.

In $B\to D^{*-}\rho^+$, the single parameter $r$ is replaced by a
matrix $\rho_{mn}$ of ratios between the magnitudes of the $\btou$ and
$\btoc$ contributions of the of three different helicity amplitudes
contributing to the decay.  
It has been shown~\cite{London:2000zi} that $\rho_{mn}$ may be obtained
using only first-order ${\cal O}(\rho_{mn})$ terms.
Not having to rely on small second-order terms 
or external input regarding amplitude ratios, this provides
a much improved, theoretically clean measurement of $\twobg$.

In this paper we generalize and extend that method to other decays
that proceed through more than one intermediate state.
Examples include $B\to D^-\rho^+$, which can interfere
with $B\to D^-\rho^+(1450)$ 
and non-resonant $B\to D^-\pi^+\pi^0$;
$B\to D^- a_1^+$, where non-resonant contributions 
are expected under the $a_1$ peak;
and $B\to D^{**-} \pi^+$, where interference between several excited
charmed mesons may be realized in the decays $D^{**-} \to D\pi$ and
$D^{**-} \to D^*\pi$, in addition to possible contributions from non-resonant
decays.

In all these cases, the interfering contributions
have overlapping yet different distributions
in relevant analysis variables. The first of these variables is
the invariant mass squared $s$ of the final state of
the resonance. The second variable $s'$ typically
describes an angular distribution 
that is fully determined by the spin of the resonance. 
In the case of $B\to D^- a_1^+$, $s'$ 
corresponds to the two variables of the Dalitz plot of the $a_1$ decay.

Our method applies equally well to modes with higher excitations, such
as $B\to D^{*-}\rho^+$, $B\to D^{*-}a_1^+$, $B\to D^{**-} \rho^+$, and
$B\to D^{**-} a_1^+$, in which $s'$ corresponds to several
angular and mass-related variables. 
In addition to the interference between several
resonances and non-resonant contributions, these decays involve
several helicity amplitudes, which are treated as different
intermediate states in our method. 

\section{Measuring \boldmath $\twobg$}

Let us consider a decay of the type described above, involving the
interference of $N$ intermediate states.  We denote the final state by
$f$ ($\fbar$) if it contains a $\overline c$ ($c$) quark.
The four decay amplitudes of interest are
\begin{eqnarray}
A(\Bz \to f) = A(\Bzb \to \fbar) &=& 
        \sum_{m=1}^N A_m g_m(s,s') e^{i\Delta_m}, 
        \nonumber\\
A(\Bz \to \fbar) &=& \sum_{m=1}^N a_m g_m(s,s') e^{i(\delta_m + \gamma)}, 
        \nonumber\\
A(\Bzb \to f) &=& \sum_{m=1}^N a_m g_m(s,s') e^{i(\delta_m - \gamma)}, 
\end{eqnarray}
where $\Delta_m$ ($\delta_m$) is the CP-conserving phase 
and $A_m$
$(a_m)$ is the magnitudes of the $\btoc$ ($\btou$) decay amplitude
proceeding via intermediate state $m$,
and $g_m(s,s')$ is a known function
of the final state variables $s$ and $s'$ that depends on the nature
of the intermediate state $m$. For example, 
for $f = D^-\pi^+\pi^0$ and $m$ being the index of the 
$D^-\rho^+$  intermediate state, 
$g_m(s,s') = R(s) s'$, where $s$ is the square of the $\pi\pi^0$ 
invariant mass, 
$R(s)$ is a Breit-Wigner function, and $s'$ is the 
cosine of the angle between the momenta of the $B$ and of one of the pions, 
calculated in the $\pi^+\pi^0$
rest frame (the ``helicity'' angle). Vector-vector intermediate
states, such as $D^{*-}\rho^+$, must be further divided into the different 
helicity amplitude, each of which has a different $s'$ dependence.

With the above equations, the time-dependent decay rates for $B^0(t)
\to f$ and $B^0(t) \to \fbar$ become
\begin{subequations}
\label{eq:rates}
\begin{eqnarray}
  \Gamma(B^0(t) \to f) = e^{-\Gamma t} \sum_{m,n} 
                \kern-2mm&\bigl[&\kern-2mm
        I_{mn} + C_{mn} \cos(\Delta mt) -S^-_{mn} \sin(\Delta mt) \bigr],  
\label{eq:bf}\\
  \Gamma(B^0(t) \to \fbar) = e^{-\Gamma t} \sum_{m,n} 
                \kern-2mm&\bigl[&\kern-2mm
        I_{mn} - C_{mn} \cos(\Delta mt) -S^+_{mn} \sin(\Delta mt) \bigr],  
\label{eq:bfbar}
\end{eqnarray}
\end{subequations}
%
where for convenience we define the symbols 
\begin{eqnarray}
I_{mn} &\equiv&\frac{1}{2} \left\{g_m g_n^* \left(A_m A_n e^{-i(\Delta_n - \Delta_m)} +
        a_m a_n e^{-i(\delta_n - \delta_m)}\right)\right\}, \nonumber\\
C_{mn} &\equiv&\frac{1}{2}  \left\{g_m g_n^* \left(A_m A_n e^{-i(\Delta_n - \Delta_m)} -
        a_m a_n e^{-i(\delta_n - \delta_m)}\right)\right\}, \nonumber\\
S^-_{mn} &\equiv& \Im\left\{ g_m g_n^* 
        A_n a_m e^{i(\delta_m - \Delta_n)} e^{i\phi} \right\}, \nonumber\\
S^+_{mn} &\equiv& \Im\left\{ g_m g_n^* 
        A_m a_n e^{-i(\delta_n - \Delta_m)} e^{i\phi} \right\}, \nonumber\\
\phi&\equiv&-(\twobg).\nonumber\\
\label{eq:defs}
\end{eqnarray}
%
The decay rates for $\Bzb$ decays are obtained from the $\Bz$ rates by
inverting the sign of the $\cos(\Delta mt)$ and $\sin(\Delta mt)$
terms. They double the statistics but do not yield additional
information.

Next, we determine the conditions under which all the unknown
parameters of Eqs.~(\ref{eq:rates}) can be obtained from the measurement,
and show that these conditions are satisfied in the typical case of
interfering Breit-Wigner resonances and a possible non-resonant
contribution.

The three terms of Eq.~(\ref{eq:bf}) are distinguishable based on
their different time dependences, thus determining their coefficients.
The relative differences between $I_{mn}$
and $C_{mn}$ are of order $(a_m a_n) / (A_m A_n) \sim r^2 \sim 10^{-4}$,
which is practically unobservable. As a result, these terms yield the
parameters $A_m$ and $\Delta_m$, while $a_m$ and $\delta_m$ are
measured from the coefficients of the $\sin(\Delta mt)$ terms, as
described later.
To study the conditions for obtaining  $A_m$ and $\Delta_m$,
we expand 
\begin{eqnarray}
\sum_{m,n} A_m A_n \left\{ g_m g_n^*  e^{-i(\Delta_n - \Delta_m)}\right\}
        &=&
        \sum_m |g_m|^2 A_m^2 \nonumber\\
        &+& 2 \sum_{m<n} \Re(g_m g_n^*) A_m A_n \cos(\Delta_n - \Delta_m) 
        \nonumber\\
        &+&  2 \sum_{m<n} \Im(g_m g_n^*) A_m A_n \sin(\Delta_n - \Delta_m) .
\label{eq:const-coef}
\end{eqnarray}
If $|g_m|^2$, $\Re(g_m g_n^*)$, and $\Im(g_m g_n^*)$ all
have different $s$ and/or $s'$ dependences, Eq.~(\ref{eq:const-coef})
yields $N^2$
unique observables, which is more than enough to determine the 
$2N - 1$ unknowns $A_m$ and $\Delta_m$ (one of the $\Delta_m$ phases is a
global phase and can be chosen arbitrarily) for $N\ge 2$.
This uniqueness condition is satisfied when all the $g_m$ are Breit-Wigner
functions, 
\begin{equation}
g_m(s) =\frac{M_m \Gamma_m}{ s - M_m^2 + i M_m \Gamma_m},
\end{equation}
even when all contributions have the same $s'$ dependence. A
non-resonant contribution $g_1 = 1$ introduces $N-1$ degenerate
relations:
\begin{equation}
\Im(g_1 g_m^*) = |g_m|^2,
\end{equation}
where $g_m \ (m>1)$ is a Breit-Wigner function.  In this case, the
number of observables is reduced to $N^2 - (N-1)$. However, a solution
still exists for $N \ge 2$, and 
this solution is unambiguous when
the non-resonant contribution is small enough relative to the resonant
contributions. In addition, most practical cases involve 
resonances with total spins different from 0, and hence $s'$ dependences
that distinguish them from a non-resonant $s$-wave contribution. This 
guarantees a unique solution of Eq.~(\ref{eq:const-coef})
in terms of $A_m$ and $\Delta_m$.

We note that these conclusions do not depend on the assumption that the
$a_m a_n$ terms in $I_{mn}$ and $C_{mn}$ are negligible.
In fact,
they apply equally well to the 
$a_m a_n \{g_m g_n^* e^{-i(\delta_n - \delta_m)}\}$
terms in Eq.~(\ref{eq:defs}).

We now show how the coefficients of the $\sin(\Delta mt)$
terms in Eq.~(\ref{eq:rates}) yield the 
values of the remaining $2N + 1$ unknowns, namely,
$a_m$, $\delta_m$, and $\phi$. The coefficients are 
\begin{eqnarray}
  \sum_{mn} S^{\mp}_{mn} &=& 
  \sum_m A_m a_m |g_m|^2 \sin(\phi \pm \delta_{mm}) \nonumber\\
  &+&  \sum_{m<n} \Im(g_m g_n^*) [\mp A_m a_n  \cos(\phi \pm \delta_{nm})
                             \pm A_n a_m  \cos(\phi \pm \delta_{mn}) ] 
        \nonumber\\
&+&  \sum_{m<n} \Re(g_m g_n^*)  [A_m a_n  \sin(\phi \pm \delta_{nm})
                             + A_n a_m  \sin(\phi \pm \delta_{mn}) ]  
\label{eq:sincoeff}
\end{eqnarray}
where $\delta_{nm} \equiv \delta_n - \Delta_m$.
If $|g_m|^2$, $\Re(g_m g_n^*)$, and $\Im(g_m g_n^*)$ are all
different, Eq.~(\ref{eq:sincoeff}) yields $N^2$ observables for
$S^-_{mn}$ and $N^2$ for $S^+_{mn}$. It is therefore possible to obtain
all the unknowns for $N \ge 2$.

\section{Discrete Ambiguities}

In the $N=1$ case, only the first line in Eq.~(\ref{eq:sincoeff}) is
non-vanishing. The measurement of $\phi$ then suffers from an
eight-fold ambiguity, due to the invariance of the observable
$\sin(\phi \pm \delta_{mn})$ under the three symmetry
operations~\cite{Silva:2002mt}:
\begin{equation}
\begin{array}{rclcl}
\Sex    &\equiv& \phi \to \delta_{mm} + \pi/2 & , & \delta_{mm} \to \phi - \pi/2, \\
\Spi    &\equiv& \phi \to \delta_{mm} + \pi   & , & \delta_{mm} \to \phi + \pi  , \\
\Spm    &\equiv& \phi \to \pi - \phi    & , & \delta_{mm} \to -\delta_{mm} .
\end{array}
\label{eq:ambig}
\end{equation}
In the typical $N>1$ case, $\Sex$ and $\Spm$ are no longer good symmetries,
since they are broken by the $\cos(\phi \pm \delta_{mn})$ terms.
Furthermore, these terms are distinguishable from the
$\sin(\phi\pm\delta_{mn})$ terms by virtue of the different $s$ and/or $s'$
dependences of $\Re(g_m g_n^*)$ and $|g_m|^2$ or $\Im(g_m g_n^*)$.
This further improves the measurement of $\phi$.

\section{Discussion and Summary}

In this paper, we have outlined the formalism for measuring
$\twobg$ with neutral $B$ meson decays involving interference between
several intermediate states.  We have shown that, 
despite involving a more complicated analysis, these decays
have distinct advantages over $B\to
D^{(*)-}\pi^+$, once our formalism is applied to their analysis,
thus enhancing the overall precision with which $2\beta+\gamma$ is known. 

First, as already noted for the special case of $B\to D^{*-} V^+$
decays~\cite{London:2000zi}, our method is sensitive to $\twobg$ using
only first order terms in the ratios $a_m / A_n$ between the $\btou$
and $\btoc$ amplitudes. By contrast, in $B\to D^{(*)-}\pi^+$,
or in the analysis of other decay modes that ignores
the contribution of multiple intermediate states, one needs
to extract $r = a_1 / A_1$ from ${\cal O}(1 - r^2)$ terms, or
rely on external measurements and incur a large theoretical
uncertainty. 
Since $r$ is as small as $1-2\%$, this advantage is realized in
our method even when the amplitude of one of the interfering
intermediate states is much greater than the others. 

Second, the $B\to D^{*-}\pi^+$ measurement is subject to an eight-fold
ambiguity, while in our method, the ambiguity is only two-fold.

Our method is not completely model-independent, since one has to
assume specific forms for the $g_m$ functions, such as a
Breit-Wigner for the resonances. However, this model dependence is
much smaller than the 30\% theoretical error estimated for $r$. Most
resonances are well understood, and their shapes can be studied with
the terms of Eqs.~(\ref{eq:const-coef}). 
In addition, the number of observables in Eq.(\ref{eq:sincoeff}) 
is greater than the number of unknowns
when there are more than two intermediate states. The additional
constraints may be used to further reduce the model dependence associated with
some $g_m$ functions. 

We emphasize that these conclusions and the formalism presented here do not
depend on a specific final state, but apply whenever enough is known
about the $g_m(s,s')$ functions for a solution to be 
obtainable, which in practice holds for a majority of the
cases.

It is interesting to note some similarities and differences between the
method we present here and methods developed for measuring $\gamma$ in
$B\to DK$.
Multi-body final state $B\to DK$ decays (such as 
$B^-\to DK^-\pi^0$~\cite{Aleksan:2002mh}, 
$B^-\to D^{**}K^-$~\cite{Sinha:2004ct}, 
$B^-\to D^*K^{*-}$~\cite{Sinha:1997zu},
or $B\to DK$ with multi-body $D$ decays~\cite{Giri:2003ty}) 
have been shown to improve the measurement of $\gamma$. This improvement is 
mostly due to the resolution of ambiguities and the ability to make 
efficient use of many $B$ and $D$ modes.
As we have shown here, similar advantages are realized by interference 
between intermediate states in the measurement of $2\beta+\gamma$ with 
multi-body $B\to D^-\pi^+$-like modes.  But in addition, these 
measurements benefit mostly from the fact that they do not depend on the 
very small $r^2$ terms. By contrast, $B\to DK$ decays are governed by 
the amplitude ratio $r_B =|A(B^-\to \overline D^0 K^-)/ A(B^-\to D^0K^-)| \sim 
10-20\%$, which is about an order of magnitude larger than $r$. 
Therefore, the sensitivity advantage brought about by interference 
between intermediate states is much greater in $B\to D^-\pi^+$ than in 
$B\to DK$.

\section{Acknowledgments}
We thank Yuval Grossman for useful comments.
The work of A.S. was supported by the U.S. Department of Energy under
contract DE-FG03-93ER40788.  The work of N.S. was supported by the
Department of Science and Technology, India.

\end{document}